\documentclass[runningheads,a4paper]{llncs}
\usepackage{latexsym}
\usepackage{multirow}
\usepackage{placeins}
\usepackage{caption}
\usepackage{subcaption}
\usepackage{float}
\usepackage{amsmath}
\usepackage{algorithm}
\usepackage{algpseudocode}
\usepackage{enumitem}
\setlist{nolistsep}
\usepackage{url}
\usepackage{graphics,graphicx}
\graphicspath{{figures/}}
\usepackage{comment}
\usepackage{color}
\usepackage{epstopdf}
\usepackage{epsfig}
\usepackage{adjustbox}
\usepackage{hhline}
\usepackage{array, caption, tabularx, ragged2e,  booktabs}

\renewcommand\textfloatsep{2pt}

\renewcommand\intextsep{2pt}

\authorrunning{Mullick et al.}

\begin{document}

\title{Public Sphere 2.0: Targeted Commenting in Online News Media}

\author{Ankan~Mullick$^1$, Sayan Ghosh$^2$, Ritam Dutt$^2$, Avijit Ghosh$^2$, Abhijnan Chakraborty$^3$\\ $^1$Microsoft, India \;		$^2$Indian Institute of Technology Kharagpur, India \\ $^3$Max Planck Institute for Software Systems, Germany\\}

\institute{}

\maketitle

\begin{abstract}
With the increase in online news consumption, to maximize advertisement revenue, news media websites try to attract and retain their readers on their sites. One of the most effective tools for reader engagement is commenting, where news readers post their views as comments against the news articles. Traditionally, it has been assumed that the comments are mostly made against the full article. In this work, we show that present commenting landscape is far from this assumption. Because the readers lack the time to go over an entire article, most of the comments are relevant to only particular sections of an article. In this paper, we build a system which can automatically classify comments against relevant sections of an article. To implement that, we develop a deep neural network based mechanism to find comments relevant to any section and a paragraph wise commenting interface to showcase them. We believe that such a data driven commenting system can help news websites to further increase reader engagement.
\end{abstract}


\vspace{-8mm}
\section{Introduction}
Recent years have witnessed a paradigm shift in the way people consume news. Online news media has become more popular than the traditional newsprint, especially to younger news readers\footnote{http://news.bbc.co.uk/2/hi/business/8542430.stm}. To further engage them, in addition to presenting news, online news platforms also allow readers to comment and share their points of view on the matter reported in stories. Irrespective of  concerns about quality of the comments, especially their language and tone, comments are considered to be the most effective tool to increase reader engagements~\cite{park2016supporting}.

Several prior works in media and communication studies have highlighted the importance of discussions in the evolution of a democratic society. In a seminal work, Habermas established the notion of `Public Sphere' where public opinion gets formed via {\it rational-critical debates}~\cite{habermas1990moral}. Ruiz {\it et al.}~\cite{ruiz2011public} argued that online news media provide a new manifestation of the public sphere -- {\it Public Sphere 2.0}, where commenting acts as the facilitator of public debates.

However, the myriad plethora of news websites today has resulted in a gradual decline of the attention span of an user to a particular news story. In an earlier work, Nielson \cite{nielsen2003usability} has noted that the readers predominately read online web pages in an F-shaped pattern i.e., two horizontal stripes in the top of the page followed by a vertical stripe along the page. This implies that the attention span of users wanes as they go through an article and most of their attention is focused on the initial paragraphs.
%
In this context, it is important to understand whether the commenting options in news websites today can felicitate discussions on the news stories and play the role of public sphere 2.0. 

To investigate this issue,  we gather articles and corresponding comments from two popular news websites -- The Guardian ({\tt theguardian.com}) and The New York Times ({\tt nytimes.com}).
We observe that a large number of comments are made targeting particular sections of an article, rather than the entire article itself. Yet, most news media websites allow their readers to comment only on the full article. In this paper, we propose to revamp the commenting UI by automatically placing the most relevant comments against each section of an article. For this, we 
develop a neural network based mechanism to map comments to particular paragraphs. Extensive evaluations show that our proposed methodology outperforms state-of-the-art baselines. Finally, we build a system which allows a reader to check for comments made against any section of an article and comment on the same. We believe that such system can help news websites in increasing reader engagement further.

\vspace{-2mm}
\section{Dataset and Motivation}
\vspace{-3mm}
In recent years, news media sites have seen huge increase in user engagement through commenting, liking, sharing etc. However, users do not spend similar time over the entire news article. 
Nielsen~\cite{nielsen2003usability} observed that, for news articles, users mostly focus on initial paragraphs or few sentences of a paragraph to consume the summary of an article, possibly due to limited time to read the whole story. 

\begin{figure}[t] 	
	\centering
	\begin{subfigure}[b]{0.45\textwidth}
		\includegraphics[width=\textwidth]{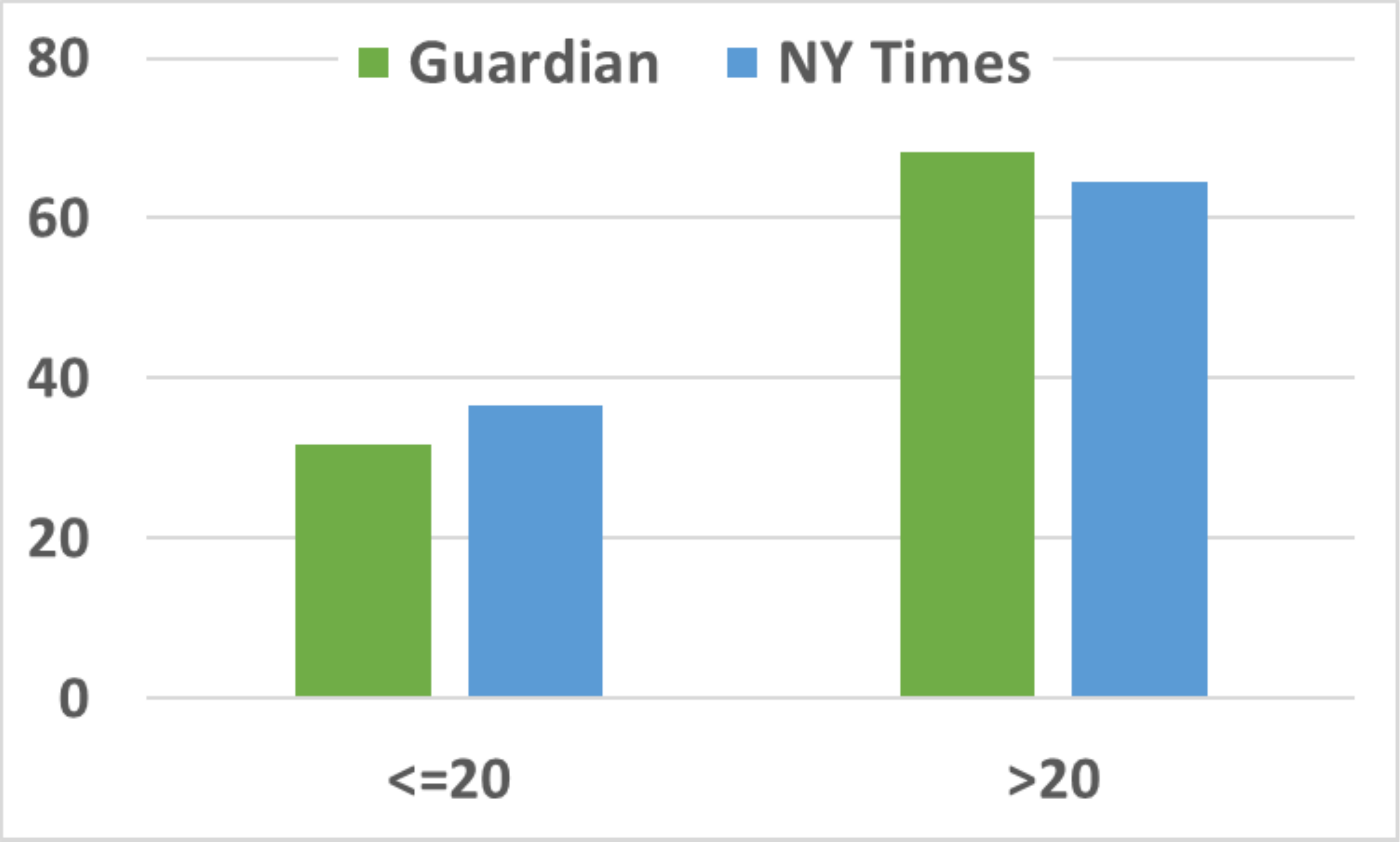}
		\caption{paragraphs vs comments}
		\label{fig:para_vs_comm}
	\end{subfigure}
     \hspace{2mm}
	\begin{subfigure}[b]{0.45\textwidth}
		\includegraphics[width=\textwidth]{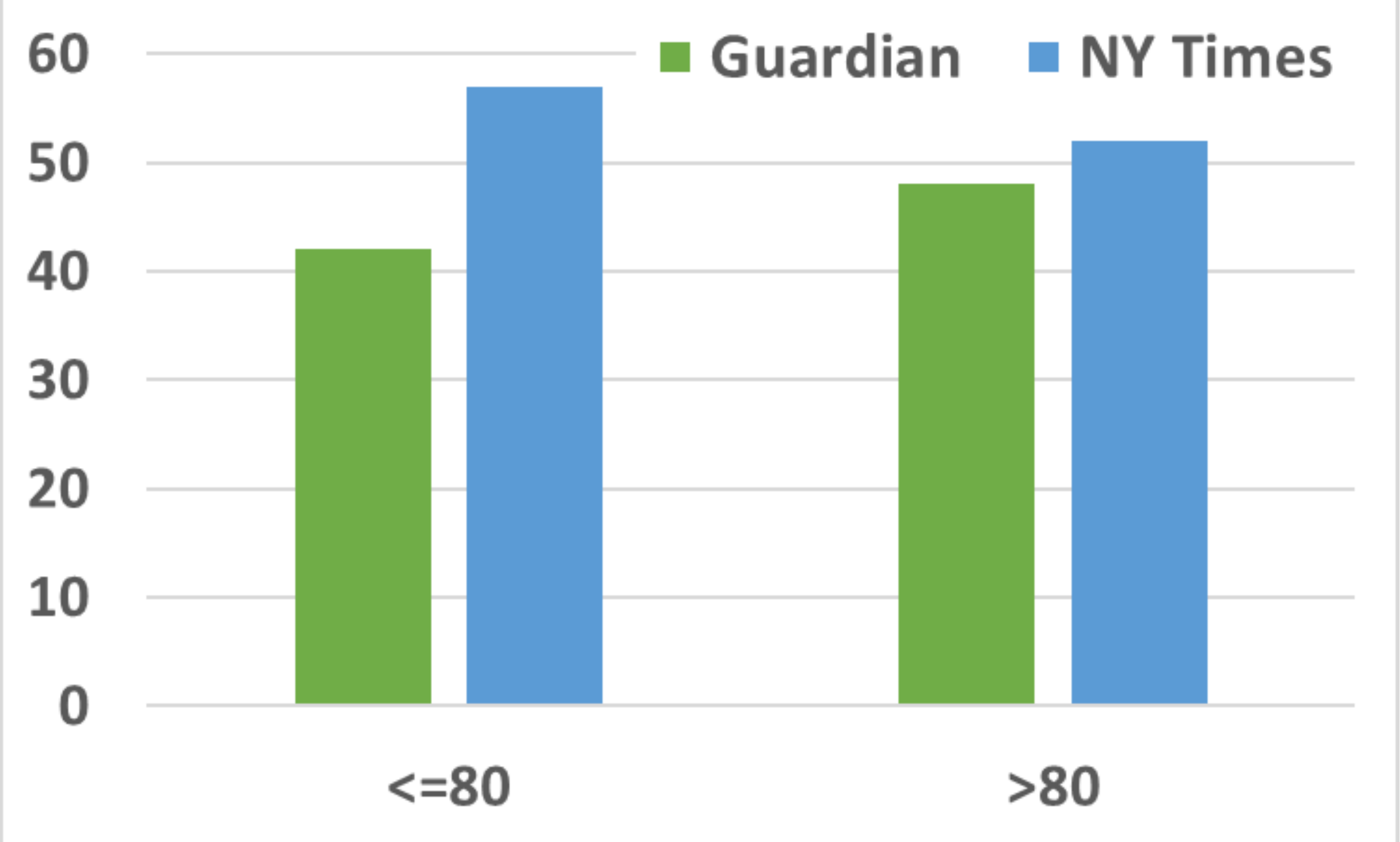}
		\caption{sentences vs comments }
		\label{fig:sentences_vs_comm}
	\end{subfigure}
     \vspace{2mm}
    \label{fig:para_sen}
    \caption{\bf Comment count varies with paragraph and sentence count.}
    \vspace{2mm}
\end{figure}

To investigate how this influences the commenting behavior, we gathered news articles from two popular news websites - `The Guardian' and `The New York Times'. In total, we collected $1,352$ Guardian and $1,020$ NYTimes news articles encompassing various topics like Business, Technology, Politics, Sports and Editorials and all comments made against these articles\footnote{https://tinyurl.com/paragraph2comment}. 

Figure \ref{fig:para_vs_comm} and \ref{fig:sentences_vs_comm} show how the number of comments varies w.r.t. the number of  paragraphs and sentences in an article (Y-axis is \% distribution). Fig \ref{fig:para_vs_comm} points out that more than $60\%$ comments are posted to the articles having more than 20 paragraphs. 
Fig \ref{fig:sentences_vs_comm} shows how comment distribution varies for 80 sentence threshold ($\sim$20 paragraphs) for two online news papers. Overall, we see that having more paragraphs in an article increases the number of comments posted against it. Thus, we can conclude that comment-paragraph relation is important. 

From the collected articles, we randomly selected $50$ articles from each media site 
for manual annotation, where two annotators were asked to give one of five possible relevance scores for a comment to a paragraph. The relevance scores are 1 (strongly irrelevant), 2(weakly irrelevant), 3(neutral), 4(weakly relevant) and 5(strongly relevant), where 
the relevance is judged by the presence and absence of common words or a common thought between the paragraph and the comment text. Both annotators provided a relevance score for each paragraph-comment pairs in all $100$ articles. Inter-annotator agreement (Cohen $\kappa$) was $0.71$.
A particular relevance score to a comment-paragraph pair was granted when both the annotators agreed. 

We observed that around $42.7\%$ of the comments (in total) were relevant to the whole article as those were not mapped to a particular paragraph. We consider a comment to be related to the entire article if the comment has a relevance score $\geq 4$ for at least 3 paragraphs or has a relevance score of $\leq 2 $ for all the paragraphs of the article.

However, approximately half of the comments ($48.9\%$ and $48.8\%$) of the Guardian and NYTimes articles are centered towards $2-3$ particular paragraphs as opposed to the entire article. 
Similar to~\cite{nielsen2003usability}, we also observe that the mean relevance of a comment decreases along the article's length. This exemplifies that more relevant comments are related to the beginning paragraphs of an article and such a trend holds true for both Guardian and NYT articles. 

Thus it is an interesting problem to find out how comments are related to individual paragraphs rather than the whole article. To automatically find out this association, 
we created the gold standard annotated datasets of 1834 and 1114 comments for `The Guardian' and `New York Times' respectively. The detailed statistics of the different annotated labels are provided in Table \ref{tab:data}. Using this data (after class balancing using the SMOTE \cite{chawla2002smote} algorithm), we design an automated approach as explained next.




\begin{table}[t]
\centering
\begin{tabular}{|c|c|c|}
\hline
Relevance Label & \% in The Guardian & \% in NY Times\\
\hline
1 & 31.05 & 40.11\\ 
\hline
2 & 19.09 & 10.23\\
\hline
3 & 17.77 & 17.50\\
\hline
4 & 19.08 & 12.29\\
\hline
5 & 13.01 & 19.87\\
\hline
\end{tabular}
\vspace{3mm}
\caption{\bf Distribution of different labels for two datasets.}
\label{table:guardian_class_dist}
\label{tab:data}
\end{table}

\vspace{-2mm}
\section{Linking comments to paragraphs}
\vspace{-1mm}
In this paper, we propose an approach to correctly identify paragraph-comment pairs and encourage users to comment towards the paragraphs, instead of only commenting on the whole article. Our proposed framework is based on deep neural networks. We have used two different neural network models - Long Short-Term Memory (LSTM) and Gated Recurrent Unit (GRU)  where inputs are paragraph and comment vectors. We have used the pre-trained 300 dimension Google News Vectors for each word and in case a pre-trained embedding for a word is not found we take it to 0 (in 300 dimension space). In order to calculate the vector for the entire paragraph and comment, we take the  average of all word vectors corresponding to each word in the paragraph and comment respectively. Deep neural network models - (i) LSTM and (ii) GRU were applied on top of the paragraph and comment vectors to get a 150 dimension vector for both paragraph and comment\footnote{After experimenting with different dimensions, results (in terms of precision, recall) were best for 150 dimension.}. Thereafter these two vectors were merged and on top of it a fully connected layer with 5 units (for five classes) and soft-max activation is applied to get the probability for each class. The proposed model is shown in Figure \ref{fig:DL_model_arch}. No explicit feature extraction, using POS Tagger or LIWC was required for these models.

\begin{table*}[t]
\centering
\scalebox{0.8}{
\begin{tabular}{|l|l|l|l|l|l|l|l|l|l|l|l|l|}
\hline
 & \multicolumn{6}{l|}{The Guardian} & \multicolumn{6}{l|}{New York Times} \\ \hline
 & \multicolumn{3}{l|}{Precision} &\multicolumn{3}{l|}{Recall} & \multicolumn{3}{l|}{Precision} &\multicolumn{3}{l|}{Recall}  \\ \hline
Model & Macro   & Micro   & Weighted   & Macro   & Micro   & Weighted   & Macro   & Micro   & Weighted   & Macro   & Micro   & Weighted     \\ \hline
NB & 46.2   & 42.6   & 61.6   & 42.6   & 42.5  & 42.6   &  33.9  & 35.2   & 60.9   & 40.9 & 35.2& 35.2 \\ \hline
DT &  42.9  & 49.6   & 53.5   & 35.6   & 50.9  & 50.9   & 36.9   & 59.2   &52.7 & 31.2  &59.1& 59.2\\ \hline
RF & 37.9   & 44.7   & 45.1   & 24.4   & 44.6  & 44.7   & 17.5   & 57.6   & 37.8   &20.2 &57.3& 57.6 \\ \hline
K-NN &  48.5  & 63.5   & 61.3   & 48.1   & 63.4  & 63.5   & 37.6   & 58.9   & 55.9   & 34.8&61.2&61.2  \\ \hline
R-SVM &  49.9  & 63.1   & 60.9   & 45.4   & 63.0  & 63.1   &  39.4  & 61.9   & 51.9 &27.3& 61.9 & 60.3  \\ \hline
AdaBoost &  38.3  & 49.3   & 48.2   & 35.7   & 49.3  & 49.2   & 29.3   & 56.1   & 48.1&28.5& 55.1   &54.6   \\ \hline
LR &  41.2  & 54.0   & 51.3   & 38.8   & 54.1  & 54.0   & 34.1   & 60.6   & 50.7   &25.8 &60.7&60.1  \\ \hline

LSTM &  64.1  & 74.4   & 74.5   &  63.6  & 74.5  &  73.3  & 56.6   & 76.8   &76.1 &57.8& 76.9  &76.8   \\ \hline
GRU &  \textbf{64.2}  & \textbf{75.3}   & \textbf{75.9}   & \textbf{63.7}   & \textbf{75.3}  &  \textbf{75.4}  &  \textbf{64.8}  & \textbf{79.1}   &\textbf{78.4}  &\textbf{64.3}& \textbf{79.3} & \textbf{79.1}  \\ \hline

\end{tabular}
}
\vspace{3mm}
\caption{\bf Performance of different models on the two datasets.}
\label{tab:results}
\end{table*}
\medskip


\vspace{-4mm}
\subsection{Baselines}
Other than neural network models, we have experimented with various traditional machine learning models  - Naive Bayes (NB), Decision Tree (DT), Random Forest (RF), K-Nearest Neighbors (K-NN), RBF Support Vector Machine (R-SVM), Logistic Regression (LR) and Adaboost. We have extracted different features for these models, which can be grouped into three different categories.

 
\noindent\textbf{POS Tag and Dependency Features:} Stanford Part-Of-Speech Tagger \cite{manning2014stanford} and Stanford dependency parser \cite{de2008stanford} were used to get different Parts-Of-Speech based features. Total 45 features were  extracted.

\noindent\textbf{LIWC Features:} Total 63 psycholinguistic features were extracted using the LIWC tool~\cite{tausczik2010psychological}.

\noindent\textbf{Others:} Uni-gram, bi-gram, tri-gram features for paragraphs and comments. 

After generating the feature matrix, dimensions were reduced using Latent Semantic Indexing (LSA) before feeding into the traditional ML-classifiers. 

\subsection{Evaluation}
\begin{figure}[t]   	
	\centering
    \begin{subfigure}[b]{0.3\textwidth}
	\includegraphics[width=\textwidth]{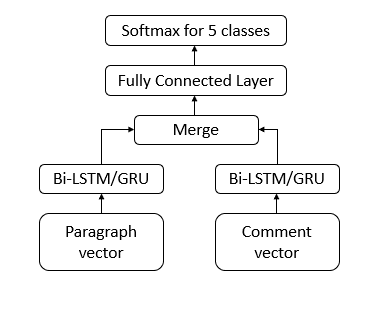}
		\caption{Architecture of the DNN model}
		\label{fig:DL_model_arch}
	\end{subfigure}
	\hspace{3mm}
	\begin{subfigure}[b]{0.5\textwidth}
		\includegraphics[width=\textwidth]{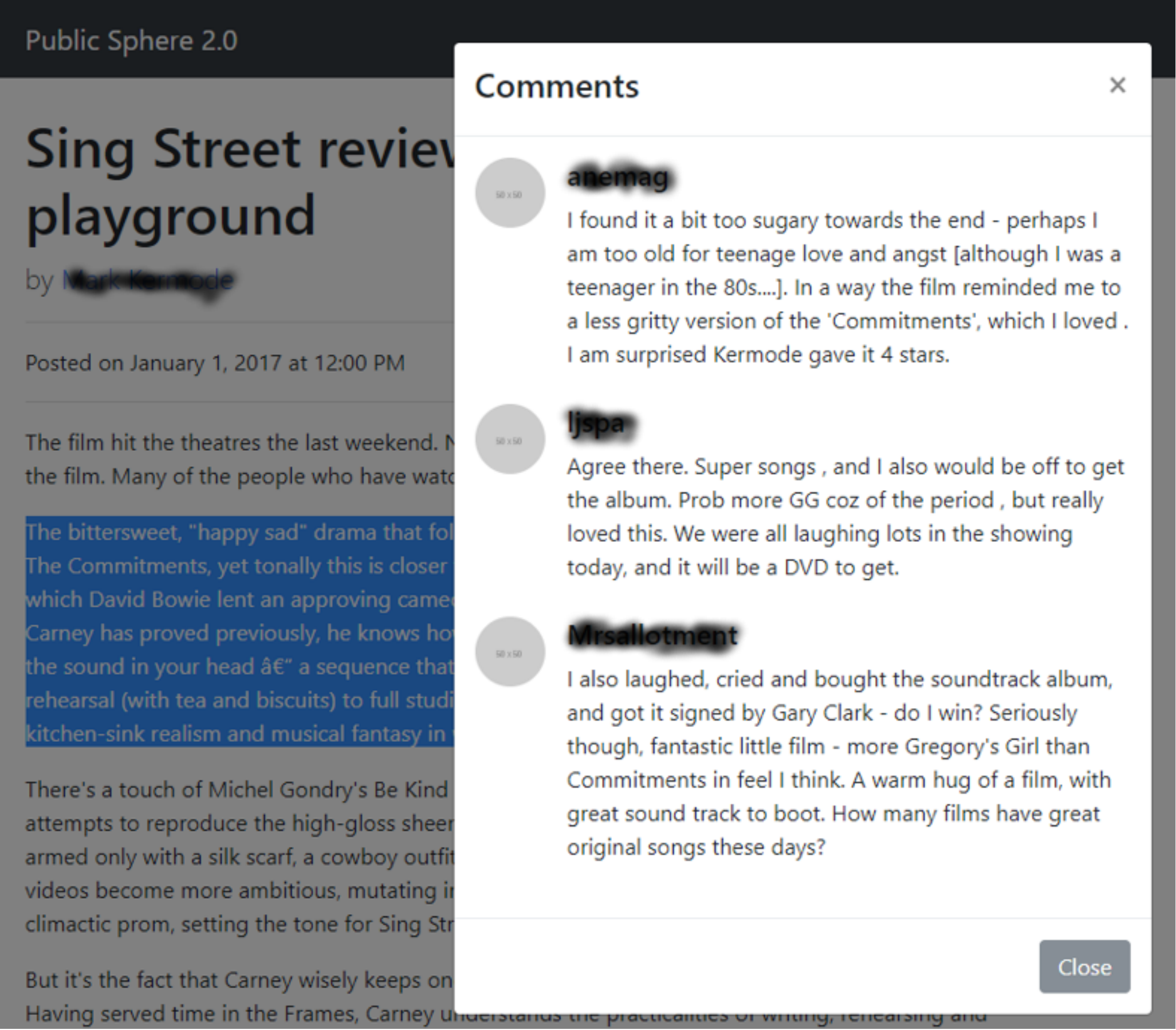}
		\caption{Snapshot of the system}
		\label{fig:system_arch}
	\end{subfigure}
	\caption{\bf Our proposed system.}
\end{figure}

 After feature extraction of the annotated datasets, various ML-classifiers were used to calculate 10-fold cross validation tests.  For the deep learning model, we have trained for 5 epochs for each step in the 10-fold cross validation.
 Results are shown in terms of Macro, Micro and Weighted averaged precision and recall for `The Guardian' and `New York Times' datasets\footnote{For ML-classifiers, we have computed precision and recall for different combination of (i) POS Tag and Dependency, (ii) LIWC and (iii) Others features but due to space constraint only the best results were shown.}. Table \ref{tab:results} shows that LSTM and GRU models outperform ML-classifier models in terms of all metrics and GRU model performs the best. Figure \ref{fig:system_arch} shows the snapshot of our model where top k (here k=3) relevant comments are highlighted when the cursor is placed around the second paragraph of a particular story.

To check the effectiveness of our system, we showed to 20 volunteers the same $10$ Guardian news stories on the original website and through our system. At the end, the volunteers were asked to rate the interface better for commenting against the articles. 17 out of 20 volunteers gave higher rating to our system interface, and the main reason they cited is the ability to see old comments and post new comments against different portions of the articles.
\section{Related Works}
\vspace{-3mm}
Here, we briefly survey the prior works on commenting in online news media.  

\noindent\textbf{Comment Ranking:} 
Hsu et al. \cite{hsu2009ranking} developed a regression model for identifying and ranking comments within a
Social Web community based on the community's expressed preferences. Dalal et al. \cite{dalal2012multi} built Hodge decomposition based rank aggregation technique to rank online comments on the social web. 

\noindent\textbf{Comment Recommendation:} Bansal et al~\cite{bansal2015content} proposed `Collaborative Correspondence Topic Models' to recommend comment-worthy blogs or news stories to a particular user (i.e., where she would be interested to leave  comments on them), where user feature profile is generated by content analysis. Shmueli et al. \cite{shmueli2012care} combined content-based approach with a collaborative-filtering approach (utilizing users' co-commenting patterns) for personalized recommendation of stories to users for discussing through comments. Agarwal et al. \cite{agarwal2011personalized} focused on personalized user preference 
based ranking of the comments in an article. 

\noindent\textbf{Comment Analysis:} Liu \cite{liu2009comment} ranked interest based news sections and articles by using a passage retrieval algorithm.  
Stroud et al. \cite{stroud2016news} analyzed demographics, attitudes and  behaviors of user population who comment on different sections. Similar analysis has also been done by Chakraborty et al~\cite{chakraborty2017tabloids,chakraborty2017makes} for social media posts. Mullick et al. \cite{mullick2017generic,mullick2018identifying} classified online comments into opinion and fact and respective subcategories. Mullick et al. \cite{mullick2016graphical} developed opinion-detection algorithm for news articles. Almgren et al.~\cite{almgren2016commenting} compared commenting, sharing, tweeting and measured user participation in them. Chakraborty et al \cite{chakraborty2017optimizing,chakraborty2019equality} utilized these different popularity signals for online news recommendations. Mullick et al.~\cite{inproceedings} experimented topic drift event and characteristics in online comments.\\

\noindent Our present work is complementary to these earlier works, where our focus is to explore paragraph oriented commenting pattern and build a model to show relevant comments to a paragraph for felicitating  more commenting.\\

\vspace{-3mm}
\section{Conclusion}
\vspace{-2mm}
To play the role of the Public Sphere, online news websites need to encourage readers to comment on their articles. In this paper, we argued for a revamp of the traditional commenting interface, and for enabling commenting on selective sections of an article. We developed a deep neural network approach to link comments to particular section. We showed that Gated Recurrent Unit (GRU) model provides best results in terms of macro and micro level precision and recall. Then, we built a basic user interface to increase user engagement in online comment sections. There are few issues to be resolved in our framework - for example, the scenario where a comment belongs to multiple paragraphs, how can a viewer select two non-consecutive paragraphs to read the respective comments and showing scores for comments. Our immediate future step is to develop an end-to-end system after resolving the issues in the model to show a user top K relevant comments (further divided into different sentiment expressed in the comments), while scrolling down the paragraphs. 
We believe such data driven selective commenting systems can bring more 
specific and targeted reader engagement for online publishing houses.

\bibliographystyle{splncs}
\bibliography{main}

\begin{thebibliography}{10}

\bibitem{park2016supporting}
Park, D., Sachar, S., Diakopoulos, N., Elmqvist, N.:
\newblock Supporting comment moderators in identifying high quality online news
  comments.
\newblock In: ACM CHI. (2016)

\bibitem{habermas1990moral}
Habermas, J.:
\newblock Moral consciousness and communicative action.
\newblock MIT press (1990)

\bibitem{ruiz2011public}
Ruiz, C., Domingo, D., Mic{\'o}, J.L., D{\'\i}az-Noci, J., Meso, K., Masip, P.:
\newblock Public sphere 2.0? the democratic qualities of citizen debates in
  online newspapers.
\newblock Intl. Journ. of Press/Politics \textbf{16}(4) (2011)

\bibitem{nielsen2003usability}
Nielsen, J.:
\newblock Usability 101: Introduction to usability (2003)

\bibitem{chawla2002smote}
Chawla, N.V., Bowyer, K.W., Hall, L.O., Kegelmeyer, W.P.:
\newblock Smote: synthetic minority over-sampling technique.
\newblock Journal of AI Research \textbf{16} (2002)

\bibitem{manning2014stanford}
Manning, C., Surdeanu, M., Bauer, J., Finkel, J., Bethard, S., McClosky, D.:
\newblock The stanford corenlp natural language processing toolkit.
\newblock In: ACL: Demo. (2014)

\bibitem{de2008stanford}
De~Marneffe, M.C., Manning, C.D.:
\newblock Stanford typed dependencies manual.
\newblock Technical report, Stanford University (2008)

\bibitem{tausczik2010psychological}
Tausczik, Y.R., Pennebaker, J.W.:
\newblock The psychological meaning of words: Liwc and computerized text
  analysis methods.
\newblock Journal of Language and Social Psychology \textbf{29}(1) (2010)

\bibitem{hsu2009ranking}
Hsu, C.F., Khabiri, E., Caverlee, J.:
\newblock Ranking comments on the social web.
\newblock In: IEEE Computational Science and Engineering. Volume~4. (2009)

\bibitem{dalal2012multi}
Dalal, O., Sengemedu, S.H., Sanyal, S.:
\newblock Multi-objective ranking of comments on web.
\newblock In: ACM WWW. (2012)

\bibitem{bansal2015content}
Bansal, T., Das, M., Bhattacharyya, C.:
\newblock Content driven user profiling for comment-worthy recommendations of
  news and blog articles.
\newblock In: ACM RecSys. (2015)

\bibitem{shmueli2012care}
Shmueli, E., Kagian, A., Koren, Y., Lempel, R.:
\newblock Care to comment?: recommendations for commenting on news stories.
\newblock In: ACM WWW. (2012)

\bibitem{agarwal2011personalized}
Agarwal, D., Chen, B.C., Pang, B.:
\newblock Personalized recommendation of user comments via factor models.
\newblock In: ACL EMNLP. (2011)

\bibitem{liu2009comment}
Liu, X.:
\newblock Comment centric news analysis for ranking.
\newblock Proceedings of the American Society for Information Science and
  Technology \textbf{46}(1) (2009)

\bibitem{stroud2016news}
Stroud, N.J., Van~Duyn, E., Peacock, C.:
\newblock News commenters and news comment readers.
\newblock Engaging News Project (2016)

\bibitem{chakraborty2017tabloids}
Chakraborty, A., Sarkar, R., Mrigen, A., Ganguly, N.:
\newblock Tabloids in the era of social media? understanding the production and
  consumption of clickbaits in twitter.
\newblock Proc. ACM Hum.-Comput. Interact. \textbf{1}(CSCW) (2017)

\bibitem{chakraborty2017makes}
Chakraborty, A., Messias, J., Benevenuto, F., Ghosh, S., Ganguly, N., Gummadi,
  K.P.:
\newblock Who makes trends? understanding demographic biases in crowdsourced
  recommendations.
\newblock In: AAAI ICWSM. (2017)

\bibitem{mullick2017generic}
Mullick, A., Maheshwari, S., Goyal, P., Ganguly, N.,  et~al.:
\newblock A generic opinion-fact classifier with application in understanding
  opinionatedness in various news section.
\newblock In: WWW Companion. (2017)

\bibitem{mullick2018identifying}
Mullick, A., Ghosh~D, S., Maheswari, S., Sahoo, S., Maity, S.K., Goyal, P.,
  et~al.:
\newblock Identifying opinion and fact subcategories from the social web.
\newblock In: ACM GROUP. (2018)

\bibitem{mullick2016graphical}
Mullick, A., Goyal, P., Ganguly, N.:
\newblock A graphical framework to detect and categorize diverse opinions from
  online news.
\newblock In: PEOPLES. (2016)

\bibitem{almgren2016commenting}
Almgren, S.M., Olsson, T.:
\newblock Commenting, sharing and tweeting news.
\newblock Nordicom Review \textbf{37}(2) (2016)  67--81

\bibitem{chakraborty2017optimizing}
Chakraborty, A., Ghosh, S., Ganguly, N., Gummadi, K.P.:
\newblock Optimizing the recency-relevancy trade-off in online news
  recommendations.
\newblock In: WWW. (2017)

\bibitem{chakraborty2019equality}
Chakraborty, A., Patro, G.K., Ganguly, N., Gummadi, K.P., Loiseau, P.:
\newblock Equality of voice: Towards fair representation in crowdsourced top-k
  recommendations.
\newblock In: ACM FAT*. (2019)

\bibitem{inproceedings}
Mullick, A., Bhandari, A., Niranjan, A., Sckhar, N., Garg, S., Bubna, R., Roy,
  M.:
\newblock Drift in online social media.
\newblock (11 2018)  302--307

\end{thebibliography}

\end{document}